\documentclass[11pt,a4paper]{article}
\usepackage[T1]{fontenc}
\usepackage[utf8]{inputenc}
\usepackage[margin=25mm]{geometry}
\usepackage{amsmath,amssymb,amsthm,enumerate,microtype}
\usepackage{newtxtext,newtxmath}
\usepackage{booktabs,tikz,float}
\usepackage[ruled,vlined,linesnumbered]{algorithm2e}
\DontPrintSemicolon
\SetKwInput{KwInput}{Input}
\SetKwInput{KwOutput}{Output}
\SetAlCapNameFnt{\small\bfseries}
\SetAlCapFnt{\small}
\usepackage[hidelinks,bookmarks=false]{hyperref}
\hypersetup{pdfauthor={Amir Belgi and Zeev Nutov},pdftitle={Constant-factor approximation algorithms for minimum-cost k-node-connectivity augmentation}}
\def\kcs{{\sc $k$-CS}}
\newcommand{\dd}{\mathrm d}
\newcommand{\uu}{\mathrm u}
\newtheorem{theorem}{Theorem}
\newtheorem{lemma}[theorem]{Lemma}
\theoremstyle{definition}
\newtheorem{definition}[theorem]{Definition}
\def\kca{{\sc $k$-CA}}
\newcommand{\bs}[1]{\mathbb{#1}}
\newcommand{\FF}{\mathcal F}
\newcommand{\PP}{\mathcal P}
\newcommand{\subs}{\subseteq}
\newcommand{\problem}[3]{\begin{center}\fbox{\begin{minipage}{0.91\linewidth}
\centering\underline{#1}\par\smallskip\raggedright
\emph{Input:} #2\par\emph{Output:} #3\end{minipage}}\end{center}}

\title{Constant-factor approximation algorithms\\for minimum-cost $k$-node-connectivity augmentation}
\author{Amir Belgi\\\small Independent Researcher, Israel\\\small\texttt{belgiamir251@gmail.com}
\and Zeev Nutov\\\small The Open University of Israel, Ra'anana, Israel\\\small\texttt{nutov@openu.ac.il}}
\date{}
\begin{document}
\maketitle
\begin{abstract}
In {\sc $k$-Connectivity Augmentation}, a $k$-connected graph must be made
$(k+1)$-connected by adding a min-cost edge set. The search for a
constant approximation for minimum-cost $k$-connected spanning subgraphs dates
back three decades. We settle the one-step augmentation case for all $n$ and $k$.
The previous best approximation ratio was
$O\left(\min\left\{\log(n-k),\log\frac{n}{n-k}\right\}\right)$ even for unit costs on a prescribed set of permitted edges.
Our approximation ratios are $13/4$ for directed graphs and $17/3$ for undirected
graphs, relative to the standard biset LP. The two uniform bounds follow from a
more general theorem on covering $k$-regular biset families.
\end{abstract}

\section{Introduction}
The search for a constant-factor approximation for the minimum-cost
$k$-connected spanning subgraph problem goes back three
decades~\cite{KhullerRaghavachari1995,Khuller1996}. Even when a $k$-connected
spanning subgraph is given for free and the goal is only to raise its
connectivity by one, no constant ratio was known for arbitrary $k$.
We settle this one-step case with approximation ratios $13/4$ for directed
graphs and $17/3$ for undirected graphs.

In 1995, Ravi and Williamson~\cite{RaviWilliamson1995}
sought a ratio of $2$ for undirected augmentation. They later showed that their
primal--dual algorithm can return a solution of cost $\Omega(k)$ times
optimum~\cite{RaviWilliamson2002}. In 2009, Nutov~\cite[Section~4]{Nutov2009OpenQuestion}
asked whether augmentation on $n$ nodes admits a constant approximation ratio or whether the
logarithmic dependence on $n/(n-k)$ is unavoidable.
G\'alvez et al.~\cite{GalvezEtAl2023} still listed the undirected case as open in 2023.

Connectivity augmentation asks how cheaply a network that tolerates $k-1$ node
failures can be made to tolerate one more. A graph on $n=|V|$ nodes is
\textbf{$k$-connected} if $n\ge k+1$ and it remains strongly connected after
the removal of any set of at most $k-1$ nodes; for undirected graphs, strong
connectivity is replaced by connectivity. We study the following problem with
nonnegative edge costs.

\problem{{\sc $k$-Connectivity Augmentation} ({\kca})}
{A $k$-connected graph $G=(V,E)$, an additional set $L$ of edges with costs, and an integer $k$.}
{A minimum-cost set $J\subs L$ such that $G+J$ is $(k+1)$-connected.}

Throughout, logarithms in asymptotic approximation bounds are understood to be
at least one. The previous general approximation bound, for both directed and
undirected graphs, was
\[
 O\left(\min\left\{\log(n-k),\log\frac{n}{n-k}\right\}\right)
 \qquad\text{\cite{Nutov2014}}.
\]
This bound is constant when $n-k$ is bounded or when $n-k=\Omega(n)$, but can
grow logarithmically in the intervening range: both terms are $\Theta(\log n)$
when $n-k=\Theta(\sqrt n)$. Nutov singled out this case
in~\cite[Section~4]{Nutov2014} and returned to the question
in~\cite{NutovSurvey2018,Nutov2022}. Our algorithms remove this parameter-dependent
logarithmic loss. They allow arbitrary nonnegative costs on a restricted set of
permitted edges.

\begin{theorem}\label{thm:augmentation}
{\kca} admits LP-relative approximation ratios $13/4$ for directed graphs
and $17/3$ for undirected graphs.
\end{theorem}

Appendix~\ref{app:parameter-bound} gives a sharper directed bound as a
function of $d=n-k$. It is below $3$ for $2\le d\le14$.

\subsection{Related work}
In {\sc $k$-Connected Subgraph} ({\kcs}), the goal is to find a minimum-cost
$k$-connected spanning subgraph. Thus {\kca} is the special case of
{\sc $(k+1)$-Connected Subgraph} in which a $k$-connected spanning subgraph
is available at zero cost. Conversely, one can build a $k$-connected spanning
subgraph by raising the connectivity one unit at a time. A uniform LP-relative
ratio $\rho$ for augmentation gives an LP-relative ratio $\rho H_k$ for {\kcs}
by this method, where $H_j=\sum_{i=1}^j1/i$~\cite{NutovSurvey2018}. In particular, Theorem~\ref{thm:augmentation} gives an $O(\log k)$ approximation for both versions of {\kcs}.

For undirected {\kcs} with arbitrary edge costs, Khuller and
Raghavachari~\cite{KhullerRaghavachari1995} asked in 1995 for a ratio of $2$.
Khuller's 1996 survey~\cite{Khuller1996} identified the existence of any
constant ratio as a main open problem. Kortsarz and Nutov~\cite{KortsarzNutov2007}
surveyed the subsequent progress. Cheriyan and V\'egh~\cite[Section~6]{CheriyanVegh2014}
obtained ratio $6$ for $n\ge k^3(k-1)+k$. Fukunaga, Nutov, and Ravi~\cite{FukunagaNutovRavi2015} extended this guarantee to $n\ge k^3$.
Nutov~\cite{Nutov2022} subsequently obtained ratio $4+\epsilon$ for $n$
sufficiently large as a function of $k$ and $\epsilon$.

The directed and undirected problems are closely related, but the parameter shift matters. 
Lando and Nutov~\cite[Corollary~2.2]{LandoNutov2009} gave a cost-preserving
reduction from directed {\kca} on $n$ nodes to undirected {\sc $(k+n)$-CA}
on $2n$ nodes. Let $\rho_{\dd}(k,n)$ and $\rho_{\uu}(k,n)$ be the best
achievable approximation ratios for directed and undirected augmentation,
respectively. This reduction and the standard bidirection reduction give
\[
 \rho_{\dd}(k,n)\leq\rho_{\uu}(k+n,2n),
 \qquad \rho_{\uu}(k,n)\leq2\rho_{\dd}(k,n).
\]
Thus uniform constant-factor approximability is equivalent for the two versions; this is not an equivalence at the same fixed $k$. Our directed ratio $13/4$ gives undirected ratio $13/2$ by bidirection alone. Covering the symmetric residual directly improves this to $17/3$.

For $n\geq k+3$, Nutov~\cite{Nutov2018} obtained ratios $H_\mu+3/2$ for directed augmentation
and $2H_\mu+1$ for undirected augmentation, with $\mu=\lfloor n/(\lceil(n-k)/2\rceil+1)\rfloor$.
These give ratios at most $3$ and $4$, respectively, when $n\geq3k-5$, improving on our uniform bounds in that range.

For undirected augmentation with $k=n-4$, G\'alvez, Hyatt-Denesik, Jabal Ameli, and Sanit\`a~\cite{GalvezEtAl2023} obtained ratios $3/2$ for weighted edges and $4/3$ for unit-cost edges, and proved APX-hardness even with unit-cost permitted edges (see also Hyatt-Denesik~\cite[Chapter~5]{HyattDenesik2024}). In undirected graphs every missing edge must be added when $k=n-2$. When $k=n-3$, the missing edges left after augmentation must form a matching, so a maximum-weight matching containing all unavailable missing edges gives an optimum. Kortsarz, Krauthgamer, and Lee~\cite{KortsarzKrauthgamerLee2004} proved APX-hardness for undirected augmentation already at $k=1$.

When every missing edge may be added at unit cost, Frank and Jord\'an~\cite{FrankJordan1995} gave
a min--max formula and a polynomial-time algorithm for adding the minimum number of directed edges
to an arbitrary digraph to reach a prescribed node-connectivity. 
For undirected augmentation by one in the unrestricted unit-cost model,
Jord\'an~\cite{Jordan1995} gave an earlier additive approximation, and
V\'egh~\cite{Vegh2011} gave an exact polynomial-time algorithm for every $k$. B\'erczi and Kobayashi~\cite{BercziKobayashi2012} also gave an exact algorithm for the target connectivity $n-3$ in this model.

Long, Mahabadi, and Tarnawski~\cite{LongMahabadiTarnawski2026} study online and offline augmentation for prescribed demand pairs in a $k$-connected undirected base graph. Their offline ratio is $O(k+\log n)$, and their hardness results concern that selected-pair model, whose optimum can be arbitrarily smaller than the cost of augmenting all pairs.

Aazami, Cheriyan, and Laekhanukit~\cite{AazamiCheriyanLaekhanukit2013} exhibited basic feasible solutions of the standard LP with all coordinates $O(1/\sqrt{k})$, obstructing the usual constant-threshold iterative-rounding analysis. Both this example and the primal--dual example of Ravi and Williamson are compatible with a constant integrality gap.

\subsection{Biset covering formulation}
Our ratios are relative to the biset LP. A \textbf{biset} on $V$ is a pair
$\bs A=(A,A^+)$ with $A\subs A^+\subs V$. We call $A$ its \textbf{inner part},
$A^+$ its \textbf{outer part}, $\partial\bs A=A^+\setminus A$ its
\textbf{boundary}, and $A^*=V\setminus A^+$ its \textbf{co-set}. Its
\textbf{co-biset} is $\bs A^*=(A^*,V\setminus A)$. A biset is \textbf{proper}
if both $A$ and $A^*$ are nonempty. All biset families considered here consist
of proper bisets. An edge \textbf{covers} $\bs A$ if it goes from $A$ to $A^*$. Write $\delta_J(\bs A)$ for the edges of $J$
that cover $\bs A$. A proper biset $\bs A$ is \textbf{tight} if
$|\partial\bs A|=k$ and $\delta_E(\bs A)=\emptyset$. Augmentation requires
$n\ge k+2$. In this range an edge set augments $G$ if and only if it covers
the family $\FF$ of tight bisets. This covering formulation and the closure
properties of tight bisets go back to Frank and Jord\'an~\cite{FrankJordan1995}.
The \textbf{cost} of $J\subs L$ is $c(J)=\sum_{e\in J}c_e$, where $c_e$
is the cost of edge $e$. Write $\delta(\bs A)=\delta_L(\bs A)$ and
$z(T)=\sum_{e\in T}z_e$ for an edge vector $z$ and $T\subs L$.
The \textbf{biset LP} has value
\begin{equation}\label{eq:augmentation-lp}
 \tau(\FF)=\min\left\{\sum_{e\in L}c_ex_e:x\ge0,\quad
 \sum_{e\in\delta_L(\bs A)}x_e\ge1\quad(\bs A\in\FF)\right\}.
\end{equation}
A \textbf{fractional cover} of $\FF$ is a feasible solution of~\eqref{eq:augmentation-lp}.
We now formulate Theorem~\ref{thm:augmentation} 
in general biset terms.

\problem{\sc Biset-Family Edge-Cover}
{A graph $(V,L)$ with nonnegative edge costs and a biset family $\FF$ on $V$.}
{A minimum-cost edge-set $J\subs L$ covering $\FF$.}

\begin{definition}
Biset \textbf{inclusion}, \textbf{intersection}, and \textbf{union} are componentwise.
Bisets \textbf{intersect} if their inner parts intersect, and \textbf{cross} if
also their co-sets intersect. A biset family is \textbf{crossing} if it contains
the intersection and union of any two crossing members. It is \textbf{intersecting} if it contains the
intersection and union of any two intersecting members. It is \textbf{$k$-regular} if
intersection and union belong to the family whenever two members intersect and
their inner union has size at most $d-1$. The \textbf{co-family} of a biset
family $\FF$ is $\FF^*=\{\bs A^*: \bs A\in\FF\}$, and $\FF$ is
\textbf{symmetric} if $\FF^*=\FF$.
\end{definition}

A \textbf{core} of $\FF$, or an
\textbf{$\FF$-core}, is an inclusion-minimal member of $\FF$. After buying $J$,
the \textbf{residual family} is $\FF^J=\{\bs A\in\FF:\delta_J(\bs A)=\emptyset\}$.

We assume a polynomial-time \textbf{pair oracle} for $\FF$. Given a set $J$ of ordered pairs on $V$ and distinct nodes $u,t$, it returns the least and greatest members of
$\FF^J$ with $u\in A$ and $t\in A^*$, or reports that there are none~\cite{Nutov2014,Nutov2018}. The pairs in $J$ may be used as tests even when they are not permitted edges.
For {\kca}, one maximum flow suffices for a fixed ordered pair.

\begin{theorem}[Covering $k$-regular biset families]\label{thm:crossing}
Let $\FF$ be a crossing biset family with $|\partial\bs A|=k$ for all
$\bs A\in\FF$ such that both $\FF$ and $\FF^*$ are $k$-regular. Then assuming
a polynomial-time pair oracle, {\sc Biset-Family Edge-Cover} admits
polynomial-time LP-relative ratio $13/4$ for directed graphs and $17/3$ for
undirected graphs if $\FF$ is symmetric. Combinatorial algorithms give ratios $7/2$ and $17/3$, respectively.
\end{theorem}

It is known that the family $\FF$ of tight bisets is crossing and that
both $\FF$ and $\FF^*$ are $k$-regular
(cf.~\cite[Proposition~1.2]{Nutov2014}). For undirected
graphs $\FF$ is symmetric, since an undirected edge covers a biset if and
only if it covers its co-biset. Theorem~\ref{thm:augmentation} therefore
follows from Theorem~\ref{thm:crossing}.

\paragraph*{Ideas and techniques.}
Put $p=\lfloor(d-1)/3\rfloor$. The initialization of~\cite{Nutov2014} buys
an edge set $J$ and assigns pairwise disjoint sets of more than $p$ nodes
to the $\FF^J$-cores with at most $p$ inner nodes. We then buy an edge set
$I$ covering every member of $\FF^J$ that contains just one $\FF^J$-core,
using the halo cover of~\cite{Nutov2014}. Thus every core of $\FF^{J\cup I}$
contains at least two $\FF^J$-cores.

Fix $\bs A,\bs B\in\FF^{J\cup I}$ and $v\in A^*\cap B^*$. We assign
each residual core contained in $\bs A$ or $\bs B$ a set of more than $p$
nodes. A core with more than $p$ inner nodes contributes its inner part.
For a smaller core we use the set assigned at initialization to one of the
$\FF^J$-cores it contains. We show that these sets are disjoint and that
their union has at most $d-1$ nodes. Since $3(p+1)>d-1$, at most two
residual cores are contained in $\bs A$ or $\bs B$.

Taking $\bs A=\bs B$ shows that each residual member contains at most two
cores. With the charging argument of~\cite{Nutov2018}, we use two orders of
the cores to obtain a directed cover of cost at most $3/2$ times the residual LP
value. Three orders give a factor of $5/3$ for a symmetric family and undirected
edges. For the sharper directed bound we use the conclusion for two bisets
to control how many chosen edges cover each biset charged in the analysis.
This gives a completion of cost at most $5/4$ times the residual LP value.

\section{The covering algorithm}\label{sec:directed}
The preparation uses directed edges. For an integer $p\ge0$, write
$\FF_{\le p}=\{\bs A\in\FF:|A|\le p\}$ and call its members \textbf{small}. The following lemma is from~\cite[Theorem~2.11 and its proof]{Nutov2014}.
For completeness, we give the proof in Appendix~\ref{sec:initialization}.

\begin{lemma}[Initialization]\label{lem:initial}
Suppose $\FF_{\le p}$ is \textbf{$p$-semi-intersecting}, meaning that it contains
the intersection of any two intersecting members $\bs A,\bs B$, and also
their union whenever $|A\cup B|\le p$. One can compute in polynomial time
an edge set $J$ of cost at most $\tau(\FF_{\le p})$
and a subfamily $\PP\subs\FF_{\le p}$ with pairwise disjoint inner parts such that
\begin{enumerate}[(i)]
\item $J$ covers every member of $\FF_{\le p}$ contained in a member of $\PP$,
and $|\delta_J(\bs A)|=1$ for every $\bs A\in\PP$.
\item For the small $\FF^J$-cores $\bs C$, the sets
\begin{equation}\label{eq:witness}
 B_C=C\cup\bigcup\{U:\bs U\in\PP,\ U\cap C\ne\emptyset\}
\end{equation}
are pairwise disjoint and each has more than $p$ nodes.
\end{enumerate}
\end{lemma}

For $p\le d/2$, two intersecting small bisets have inner union of size at most
$2p-1\le d-1$. Thus $k$-regularity puts their intersection in $\FF_{\le p}$
and their union there whenever it is small. The same holds in every residual
family, since an edge covering the intersection or union covers at least one
of the two bisets.

We use $p=\lfloor(d-1)/3\rfloor$. Three small inner parts then have at most
$d-1$ nodes in their union, whereas three disjoint sets of more than $p$ nodes
have more than $d-1$ nodes. The first inequality lets us uncross a small biset
with two members of $\PP$. The second will bound the number of surviving cores.
For $d=2,3$ the small family is empty and $J=\PP=\emptyset$.
As with the area functions in the $(4+\epsilon)$-approximation
for $k$-connected subgraphs~\cite{Nutov2022}, we cover less to obtain more structure.
Appendix~\ref{sec:initialization} illustrates what changes at threshold $d/2$.

The \textbf{halo family} of an $\FF$-core $\bs C$ consists of the members of
$\FF$ containing $\bs C$ and no other $\FF$-core. For any finite biset family
$\FF$, covering all its halos leaves every residual member containing at least two
$\FF$-cores. Indeed, every member contains a core, and every member containing just
one belongs to its halo. Kortsarz and Nutov~\cite{KortsarzNutov2005} used this
observation. The cost bound in the following lemma uses the argument of
Fakcharoenphol and Laekhanukit~\cite[Section~4.3]{FakcharoenpholLaekhanukit2012},
who charged each edge to at most one halo. The formulation for crossing biset
families follows from~\cite[Lemma~2.3 and Corollary~2.6]{Nutov2014}.

\begin{lemma}[Halo cover]\label{lem:halo}
Let $\FF$ be a crossing family, and let $\mathcal H$ be the union of
the halo families of any collection of $\FF$-cores. One can compute in
polynomial time a directed cover $I$ of $\mathcal H$ of cost
$c(I)=\tau(\mathcal H)\le\tau(\FF)$.
\end{lemma}
\vspace{-\smallskipamount}

\noindent\emph{Proof sketch.} In a crossing family, the members covered by a directed edge
have a least and a greatest member whenever there are any. They all contain
its tail in their inner parts and its head in their co-sets, so they are
closed under intersection and union. If the edge covers a halo member, the
least member it covers belongs to the same halo. Thus the edge sets available
to distinct halos are disjoint. Each halo has an
intersecting co-family~\cite[Lemma~2.5]{Nutov2014}.
Section~\ref{sec:implementation} gives the polynomial-time implementation
that uses the pair oracle to cover each halo at its LP value. These values
sum to $\tau(\mathcal H)\le\tau(\FF)$.\hfill$\square$

\medskip

A family $\FF$ has the \textbf{local two-core property} if, for any
$\bs A,\bs B\in\FF$ with $A^*\cap B^*\ne\emptyset$, at most two
$\FF$-cores $\bs C$ satisfy $\bs C\subs\bs A$ or $\bs C\subs\bs B$.

\begin{lemma}[Residual cores]\label{lem:two-cores}
Let $\FF$ be as in Theorem~\ref{thm:crossing}, and suppose $J,\PP$ satisfy the
conclusions of Lemma~\ref{lem:initial} for $p=\lfloor(d-1)/3\rfloor$.
Let $I$ cover all halo families of $\FF^J$.
Then $\FF^{J\cup I}$ has the local two-core property.
\end{lemma}

Taking $\bs A=\bs B$ shows that every residual member contains at most two cores.
This weaker conclusion suffices for the following combinatorial completions.

\begin{lemma}[Core-order completion]\label{thm:completion}
Let $\FF$ be a crossing family in which every member contains at most two
$\FF$-cores. There is a combinatorial polynomial-time algorithm that computes
an edge cover of $\FF$ of cost at most
\begin{enumerate}[(i)]
\item $\tfrac32\tau(\FF)$ for directed edges,
\item $\tfrac53\tau(\FF)$ for undirected edges when $\FF$ is symmetric.
\end{enumerate}
\end{lemma}

The full conclusion of Lemma~\ref{lem:two-cores} gives a sharper directed completion.

\begin{lemma}[Rounding a directed fractional cover]\label{sharp}
Let $\FF$ be a crossing family with the local two-core property.
Then any fractional directed
cover $x$ can be rounded in polynomial time to a cover of cost at most
$\tfrac54 c^\top x$.
\end{lemma}

\medskip
\begin{algorithm}[H]
\caption{Directed cover of $\FF$}\label{alg:augmentation}
\KwInput{A family $\FF$ as in Theorem~\ref{thm:crossing} and permitted directed edges with costs.}
Apply Lemma~\ref{lem:initial} with $p=\lfloor(d-1)/3\rfloor$ to obtain $J,\PP$\;
Buy a minimum-cost cover $I$ of all halos of $\FF^J$ by Lemma~\ref{lem:halo}\;
Round an optimal fractional cover of $\FF^{J\cup I}$ by Lemma~\ref{sharp} to obtain $Y$\;
\Return{$J\cup I\cup Y$}\;
\end{algorithm}
\medskip

For the combinatorial directed algorithm, obtain $Y$ in step~3 by applying
Lemma~\ref{thm:completion}(i) to $\FF^{J\cup I}$. For undirected edges, run
steps~1 and~2 on the bidirected graph. Then buy the underlying edges of
$J\cup I$, install both orientations of each, and apply
Lemma~\ref{thm:completion}(ii) to the resulting symmetric residual family.

\subsection{Confinement of residual cores (Lemma~\ref{lem:two-cores})}\label{sec:two-cores}
Fix $\bs A,\bs B\in\FF^{J\cup I}$ and $v\in A^*\cap B^*$. To each
residual core contained in either member we will assign at least $p+1$ nodes. These
sets are pairwise disjoint and lie in a set of at most $d-1$ nodes.
Since $3(p+1)>d-1$, at most two cores fit.

Each residual family inherits crossing and
$k$-regularity: an edge covering the intersection or union of two bisets
covers at least one of them. Reversing the purchased edges gives the same
observation for the co-family. Every member satisfies $|A|+|A^*|=d$.
Moreover,
\begin{equation}\label{eq:inner-containment}
 A\subs B\quad\Longrightarrow\quad \bs A\subs\bs B\qquad(\bs A,\bs B\in\FF).
\end{equation}
Indeed, $A\cap B=A\ne\emptyset$ and $|A\cup B|=|B|\le d-1$, so $\bs A\cap\bs B\in\FF$. Its inner part
is $A$ and its outer part is contained in $A^+$. Equal boundary sizes give
$\bs A\cap\bs B=\bs A$.

The \textbf{common co-set bound} is
\begin{equation}\label{eq:common-coset}
 M^*\cap N^*\ne\emptyset\quad\Longrightarrow\quad |M\cup N|\le d-1
 \qquad(\bs M,\bs N\in\FF).
\end{equation}
Fix $v\in M^*\cap N^*$. If $|M\cup N|\ge d$, then $|M|+|N|\ge d$, and hence
$|M^*\cup N^*|\le 2d-|M|-|N|-1\le d-1$.
By $k$-regularity of $\FF^*$, $\bs M\cup\bs N$ belongs to $\FF$.
Its co-set contains $v$, so its inner part has at most $d-1$ nodes, a contradiction.

\emph{Disjointness.} The residual cores $\bs K$ with $v\in K^*$ have disjoint inner
parts. If two of them intersected, they would cross, and their intersection
would be a residual member properly contained in both. For each such core put
\[
 \widehat K=K\cup\bigcup\{U:\bs U\in\PP,\ U\cap K\ne\emptyset,\ v\in U^*\}.
\]
Every $\bs U$ in this union crosses $\bs K$. Their intersection belongs
to $\FF_{\le p}$ and lies in $\bs U$. By Lemma~\ref{lem:initial}(i), $J$
covers this intersection and misses $\bs K$. Hence the unique edge of $J$
covering $\bs U$ has its tail in every such $K$ that $U$ meets. The cores
are disjoint, so $U$ meets at most one. The inner parts in $\PP$ are also
disjoint, proving that the enlarged sets $\widehat K$ are pairwise disjoint.

\emph{Size.} We claim $|\widehat K|>p$. It suffices to consider $|K|\le p$.
Suppose two distinct members $\bs U,\bs W\in\PP$ meet $\bs K$ but omit $v$
from their co-sets.
By $k$-regularity, $\bs K\cup\bs U$ and $\bs K\cup\bs W$ belong to $\FF$.
They intersect in their inner parts, and their inner union has at most
$3p\le d-1$ nodes, so their intersection also belongs to $\FF$. Its inner
part is $K$, since $U\cap W=\emptyset$, while its outer part contains both
$K^+$ and $v\notin K^+$. Its boundary therefore has more than $k$ nodes,
a contradiction. Thus at most one member of $\PP$ meeting $K$ omits $v$
from its co-set.
Since $I$ covers all halos of $\FF^J$, $\bs K$ contains at least two
$\FF^J$-cores, both small. If the omitted member met two such cores, its nonempty inner part would
lie in both sets $B_C$ of~\eqref{eq:witness}, contradicting their disjointness. For at least one of these old cores $\bs C$ we therefore
have $B_C\subs\widehat K$. Lemma~\ref{lem:initial}(ii) gives $|\widehat K|>p$.

\emph{Confinement.} Extend $\bs A,\bs B$ to maximal members $\bs M,\bs N$
of $\FF$ with $v$ in their co-sets. For $\bs K\subs\bs A$, each $\bs U$ used
in $\widehat K$ crosses $\bs M$, and $\bs U\cup\bs M$ still has $v$ in its
co-set. Maximality gives $\bs U\subs\bs M$, so $\widehat K\subs M$.
Symmetrically, $\widehat K\subs N$ for $\bs K\subs\bs B$.

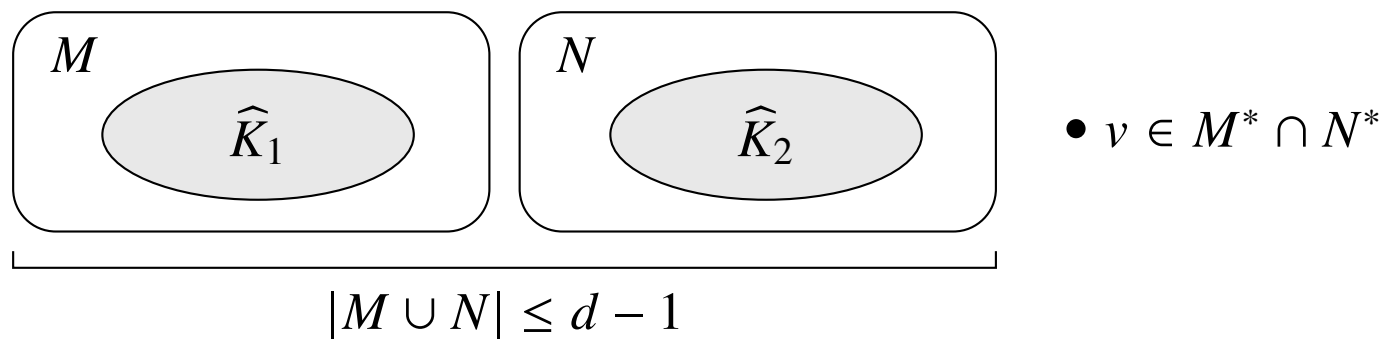
\begin{figure}[ht]
\centering
\begin{tikzpicture}[x=1cm,y=1cm,every node/.style={font=\small}]
  \draw[rounded corners=8pt] (0,0) rectangle (3.15,1.45);
  \draw[rounded corners=8pt] (3.35,0) rectangle (6.5,1.45);
  \node[anchor=north west] at (.10,1.42) {$M$};
  \node[anchor=north west] at (3.45,1.42) {$N$};
  \draw[fill=black!9] (1.62,.64) ellipse (1.03 and .43);
  \draw[fill=black!9] (4.98,.64) ellipse (1.03 and .43);
  \node at (1.62,.64) {$\widehat K_1$};
  \node at (4.98,.64) {$\widehat K_2$};
  \fill (7.03,.68) circle (1.8pt);
  \node[anchor=west] at (7.08,.68) {$v\in M^*\cap N^*$};
  \draw (0,-.13)--(0,-.24)--(6.5,-.24)--(6.5,-.13);
  \node[anchor=north] at (3.25,-.26) {$|M\cup N|\le d-1$};
\end{tikzpicture}
\caption{Packing at a fixed co-set node $v$. Each shaded set has more than
$(d-1)/3$ nodes. Two are shown in distinct maximal members; the same count
applies if both lie in one member.}\label{fig:confinement}
\end{figure}

The enlarged sets for the cores in $\bs A$ or $\bs B$ are pairwise disjoint.
Each has more than $(d-1)/3$ nodes, and all lie in $M\cup N$, which has
at most $d-1$ nodes by~\eqref{eq:common-coset}. Hence there are at most two
(Figure~\ref{fig:confinement}).

\section{Completion by ordering the cores (Lemma~\ref{thm:completion})}\label{sec:dir-finish}
Fix the cores of $\FF$ throughout. For a core $\bs C$, write
$\FF(C)=\{\bs A\in\FF:\bs C\subs\bs A\}$. Two members of $\FF(C)$
whose co-sets intersect cross because both contain $\bs C$. Their
intersection and union also contain $\bs C$. Thus $\FF(C)$ has an
intersecting co-family, as does the subfamily that remains after any purchase.

\paragraph*{Directed edges.}
Choose any order of the cores. Starting with no edges, process the cores
in this order. At each $\bs C$, buy a minimum-cost directed cover of the
still-uncovered members of $\FF(C)$. Every member of $\FF$ contains a core,
so the resulting edge set covers $\FF$. Start again with no edges, process
the cores in reverse order, and return the cheaper of the two covers.

We use the charging argument of~\cite[Lemma~6]{Nutov2018}. Fix a fractional
cover $x$. At each core charge $c_ex_e$ for every edge $e$ covering a member
still to be covered at that core. The restriction of $x$ to these edges is
feasible, so exact directed covering bounds the purchase by the total charge.
For each edge $e$ covering some member, choose a core $\bs C_e$ contained in
the least member that $e$ covers. Every member covered by $e$ contains
$\bs C_e$, so $e$ receives no further charge after $\bs C_e$ is processed.
Every core at which $e$ is charged lies in the greatest member that $e$
covers. By hypothesis this member contains at most one core besides $\bs C_e$.
That other core, if present, precedes $\bs C_e$ in exactly one of the two
orders. Thus $e$ is charged at most three times in all, and the cheaper
cover costs at most $\tfrac32\sum_e c_ex_e$.

\paragraph*{Undirected edges.}\label{sec:undirected}
Assume $\FF$ is symmetric. Each edge covering a member has two distinguished cores, one for each orientation. Processing
the first covers every remaining member the edge can cover. Each orientation
can incur one additional charge if a third core precedes both distinguished
cores. In a uniformly random order this happens with probability $1/3$,
giving $1+1/3+1/3=5/3$. We realize the same bound with three orders. Join two cores $\bs C,\bs D$ if a member of $\FF$ contains $\bs C$ and its co-biset contains $\bs D$. Equivalently, $\bs D\subs\bs C^*$, since $\bs C$ itself then separates the pair. The neighbours of $\bs C$ are precisely the cores in $\bs C^*$. As $\bs C^*\in\FF$, the core graph has maximum degree two. Greedily colour it with three colours.

Choose an arbitrary list $L_i$ for each colour $i$. Writing $\overleftarrow L_i$ for its reverse, use the three orders
\[
 L_1\,\overleftarrow L_2\,L_3,\qquad
 L_2\,\overleftarrow L_3\,L_1,\qquad
 L_3\,\overleftarrow L_1\,L_2 .
\]
Start each trial with no edges. At each core $\bs C$, compute a minimum-cost directed cover of the still-uncovered members of $\FF(C)$ on the bidirected graph. Buy the underlying undirected edges and install both orientations before processing the next core. Every trial covers $\FF$. Return the cheapest.

If $\bs C,\bs D$ have different colours, any third core $\bs E$ precedes both in at most one order. If $\bs E$ has the third colour, its colour class must come first. If $\bs E$ has the colour of $\bs C$ or $\bs D$, their common block must precede the other endpoint's block. It can then occur first or second, and its internal orders in these two positions are opposite.

Fix a fractional edge cover $x$ and an edge $e=uv$ covering some member of $\FF$. The members covered by $uv$ have a least and a greatest member. Fix a core $\bs C$ in the least. By symmetry the members covered by $vu$ are their co-bisets; fix a core $\bs D$ in the least of these bisets. Every member covered by $uv$ contains $\bs C$ and its co-biset contains $\bs D$. Thus $\bs C,\bs D$ are adjacent and have different colours.

At each core, for every edge assign its fractional value to each orientation that covers a member of the subfamily being processed. This gives a feasible fractional directed cover of that subfamily. Each such orientation therefore contributes $c_ex_e$ to the bound for that core. The first of $\bs C,\bs D$ to be processed covers all remaining bisets that $e$ can cover, by symmetry. If $\bs C$ comes first, only $uv$ can be charged there, since every member covered by $vu$ has $C$ in its co-set. The case in which $\bs D$ comes first is symmetric. For $uv$, every charging core lies in the greatest member covered by $uv$. Besides $\bs C$ there is at most one other. It differs from $\bs D$ as well, since the co-biset of that greatest member contains $\bs D$. It can charge only before both $\bs C,\bs D$. It therefore contributes at most once across the three orders. The same count holds for $vu$, even if the additional core is the same for both orientations. The three trials together charge $e$ at most $3+1+1=5$ times. Summing over edges gives cost at most $5\sum_e c_ex_e/3$ for the cheapest trial.

\paragraph*{Proof of Theorem~\ref{thm:crossing} (directed ratio $7/2$).}
By Lemmas~\ref{lem:initial} and~\ref{lem:halo},
$c(J)+c(I)\le\tau(\FF_{\le p})+\tau(\FF^J)\le2\tau(\FF)$.
Taking $\bs A=\bs B$ in Lemma~\ref{lem:two-cores}, every member of
$\FF^{J\cup I}$ contains at most two cores. By Lemma~\ref{thm:completion}, the
two-order completion costs at most $\tfrac32\tau(\FF^{J\cup I})\le\tfrac32\tau(\FF)$, giving
$\tfrac72\tau(\FF)$ in total.\hfill$\square$

\paragraph*{Proof of Theorem~\ref{thm:crossing} (undirected ratio $17/3$).}
In the bidirected preparation described after Algorithm~\ref{alg:augmentation},
copying an undirected fractional cover to both orientations doubles its cost.
Thus Lemmas~\ref{lem:initial} and~\ref{lem:halo} bound the cost of the two
purchases by $4\tau(\FF)$ in total.
The resulting family is symmetric. It is exactly $\FF^{J\cup I'}$, where $I'$
consists of $I$ and the reversals of all edges in $J\cup I$. Since $I'$
still covers all halos of $\FF^J$, Lemma~\ref{lem:two-cores} applies with
the original $J,\PP$ and with $I'$. Taking $\bs A=\bs B$, each remaining
member contains at most two residual cores. The undirected part of Lemma~\ref{thm:completion} gives a completion of cost at most $5\tau(\FF)/3$. The total is
$(4+5/3)\tau(\FF)=17\tau(\FF)/3$.\hfill$\square$

\section{Rounding a directed fractional cover (Lemma~\ref{sharp})}\label{quarter}
Let $\FF$ and $x$ be as in Lemma~\ref{sharp}. Fix the cores of $\FF$.
A member is \textbf{private} if it contains exactly one core and
\textbf{paired} if it contains two. For each core $\bs D$, let $\Lambda_D$
be the permitted edges covering a member of its halo. By the proof of
Lemma~\ref{lem:halo}, the least member covered by an edge of $\Lambda_D$
belongs to that halo. Thus the sets $\Lambda_D$ are pairwise disjoint,
and every member covered by an edge of $\Lambda_D$ contains $\bs D$.
If $\bs U$ is paired and $W^*\cap U^*\ne\emptyset$ for $\bs W\in\FF$,
the local two-core property implies that every core in $\bs W$ is one of
the two in $\bs U$.

\emph{Where the factor $5/4$ comes from.} We price some paired members
$\bs U$. Fix one, and write $y=y_U=x(\delta(\bs U))\ge1$. The two sets
$\Lambda_D$ corresponding to its cores contribute $a,b$ to this sum, with
$a+b\le y$. Suppose the two halo covers have $s$ and $t$ edges covering
$\bs U$, respectively. Let $m$ count the edges covering $\bs U$ after both
deletion passes. We show that $m=1$ if $st=0$ and $m\le s+t$ otherwise.
Hence $m\le1+st$ and $m\le1+s+t$, since
$s+t\le1+st$ when both counts are positive. Averaging these bounds with weights $1/y$ and $1-1/y$
gives $m\le g_y(s,t)$, where
\[
 g_y(a,b)=1+\left(1-\frac1y\right)(a+b)+\frac{ab}{y}.
\]
For the fractional loads, $a+b\le y$ gives
$g_y(a,b)\le y+ab/y\le5y/4$. We use this function to choose the halo covers.

\paragraph*{Claim 1. Pricing.}
Start with reduced costs $c$. Call an edge \textbf{free} when its reduced
cost is zero. While some paired biset is missed by the free edges, choose an inclusion-maximal such biset $\bs U$.
Subtract the least reduced cost $\pi_U$ of an edge in $\delta(\bs U)$ from
every edge in $\delta(\bs U)$. Write
$r_e=c_e-\sum_{U:e\in\delta(\bs U)}\pi_U$ and $Z=\{e:r_e=0\}$ at termination.
Thus $r\ge0$ and $c^\top x=r^\top x+\sum_U\pi_U y_U$. Two paired members with
intersecting co-sets contain the same two cores, so they cross, and their
intersection and union are paired. If a later priced biset $\bs T$ has $T^*\cap U^*\ne\emptyset$, their
union was therefore paired and was uncovered when $\bs U$ was chosen. Maximality gives $\bs T\subs\bs U$. Thus two priced bisets with intersecting co-sets are comparable. At termination $Z$ covers all paired members.

\emph{An earlier free edge.} Fix a priced $\bs U$ and a member
$\bs W$ whose co-set meets $U^*$ but does not contain it.
Then $\bs W$ and $\bs U$ share a core and cross. Their union is paired and
strictly contains $\bs U$. By maximality, some edge already free when
$\bs U$ was chosen covered the union and missed $\bs U$.
Its tail belongs to $W\setminus U$ and its head to $W^*\cap U^*$. In
particular $\bs W$ is covered by an edge of
\begin{equation}\label{internal}
 Z_U=\{uv\in Z:u\notin U,\ v\in U^*\}.
\end{equation}
This edge was already free just before $\bs U$ was priced, when every edge
covering $\bs U$ still had positive reduced cost.

\paragraph*{Claim 2. Rounding the halo covers.}
For each priced $\bs U$ with cores $\bs D_1,\bs D_2$, put
$a_U(z)=z(\delta(\bs U)\cap\Lambda_{D_1})$ and
$b_U(z)=z(\delta(\bs U)\cap\Lambda_{D_2})$. Keep $y_U=x(\delta(\bs U))$
fixed and consider
\[
 \Phi(z)=r^\top z+\sum_U\pi_U\,g_{y_U}\bigl(a_U(z),b_U(z)\bigr).
\]
Start with $z=x$ and set $z_e=0$ outside $\bigcup_D\Lambda_D$. This leaves
$a_U(z),b_U(z)$ unchanged for every priced $\bs U$ and cannot increase $r^\top z$.
Since the two loads at each priced biset come from distinct edge sets,
$\Phi$ is affine on each $\Lambda_D$ when the other domains are fixed,
with nonnegative coefficients.
The restriction of $x$ to $\Lambda_D$ covers the halo of $\bs D$.
Process each core once, replacing $z$ on $\Lambda_D$ by an inclusion-minimal
minimum-cost cover $H_D$ of its halo at these coefficients. Exact directed
halo covering gives this replacement without increasing $\Phi$.
For $H=\bigcup_D H_D$ write $\Phi(H)=\Phi(\mathbf1_H)$.
Then $\Phi(H)\le\Phi(x)$. For each priced $\bs U$, let $s_U,t_U$ count
the edges of $H_{D_1},H_{D_2}$ covering $\bs U$, respectively. Thus
$s_U=a_U(\mathbf1_H)$, $t_U=b_U(\mathbf1_H)$, and $s_U+t_U=|\delta_H(\bs U)|$.

\paragraph*{A witness observation.}
Let $Y'$ cover a crossing family, and let $\bs U,\bs W_1,\bs W_2$ be members
with $W_1\cap W_2\ne\emptyset$ and $U^*\subs W_1^*\cap W_2^*$.
Suppose $e_i$ is the unique edge of $Y'$ covering $\bs W_i$, and both
$e_1,e_2$ cover $\bs U$. Then $e_1=e_2$. Indeed, the witnesses cross,
so $Y'$ covers their intersection. An edge covering it covers at least one
witness, hence is $e_1$ or $e_2$. Its tail lies in both inner parts and
its head lies in $U^*$, so it covers both witnesses. By uniqueness,
$e_1=e_2$.

\paragraph*{Claim 3. First deletion pass.}
Start with $H\cup Z$, which covers every member. Keep $H$ and test
the edges of $Z\setminus H$ in reverse order of when their reduced costs
first reached zero, breaking ties arbitrarily. Delete an edge whenever
the remaining edges still cover every member. Let $Z'$ consist of
the retained edges outside $H$. Then $r(Z')=0$, and we claim that
\begin{equation}\label{q}
 |\delta_{Z'}(\bs U)|=\mathbf1_{\{\delta_H(\bs U)=\varnothing\}}\qquad
 \text{for every priced }\bs U .
\end{equation}

Fix a priced $\bs U$ and suppose $e\in\delta_{Z'}(\bs U)$. When $e$ was
tested, it alone covered some witness $\bs W$. Only deletions follow in
this pass, so $e$ still covers $\bs W$ alone in $H\cup Z'$. The witness
is paired because $H$ covers every private member. Its co-set contains
$U^*$. Otherwise the argument preceding~\eqref{internal} gives an edge
covering $\bs W$ that became free before $e$. That edge was present when $e$ was tested, either fixed in $H$
or awaiting its test in reverse order.

If $h\in H$ also covers $\bs U$, take a private $\bs X$ covered in $H$ by
$h$ alone. Such a member exists by the minimality of the halo cover
containing $h$ and the disjointness of the sets $\Lambda_D$. Its co-set meets $W^*$ at the head
of $h$, so $\bs X$ and the paired $\bs W$ share a core and cross. Their
intersection is private and covered by $H$. An edge of $H$ covering it
covers $\bs X$ or $\bs W$. Since $H$ misses $\bs W$, this edge is $h$.
Now $h$ has its tail in $W$ and its head in $U^*\subs W^*$, a contradiction.

If $H$ misses $\bs U$, at least one edge of $Z'$ covers it. For any
$e_1,e_2\in\delta_{Z'}(\bs U)$, their witnesses are paired, contain the
same two cores as $\bs U$, and have co-sets containing $U^*$. The witness
observation gives $e_1=e_2$. This proves~\eqref{q}.

\paragraph*{Claim 4. Second deletion pass.}
Start with $Y=H\cup Z'$. Among the priced bisets $\bs G$ with $s_Gt_G=0$
and $|\delta_{H\cup Z'}(\bs G)|\ge2$, select the inclusion-minimal ones.
Their co-sets are pairwise disjoint by comparability. For each selected
$\bs G$, all edges of $\delta_{H\cup Z'}(\bs G)$ belong to a single halo
cover $H_D$ by~\eqref{q} and $s_Gt_G=0$.

At each selected $\bs G$, fix the current edges with heads outside $G^*$,
and add $Z_G$ as defined in~\eqref{internal}. Test all edges with heads in
$G^*$ for deletion, preserving coverage. Test those in $H\setminus Z$ first,
then those in $Z$ in reverse order of becoming free. Different selected
bisets change disjoint sets of heads, and only free edges are added. Thus
$r(Y)\le r(H)$. We claim that on termination, for every priced $\bs U$,
\begin{equation}\label{counts}
 |\delta_Y(\bs U)|=1\quad\text{if }s_Ut_U=0,\qquad
 |\delta_Y(\bs U)|\le s_U+t_U\quad\text{otherwise}.
\end{equation}

\emph{Common core.} Just before $Z_G$ is added,
$\delta_Y(\bs G)=\delta_{H\cup Z'}(\bs G)\subs H_D$, since earlier tests
changed only edges with heads outside $G^*$.
Every member $\bs W$ missed by the fixed edges contains $\bs D$. Indeed,
a current edge covering $\bs W$ has its head in $G^*$, so $\bs W$ shares a
core with the paired $\bs G$ and crosses it. Every current edge covering
their intersection also covers $\bs G$: otherwise its head would lie in
$W^*\setminus G^*$, making it a fixed edge covering $\bs W$.
If $\bs W$ omitted $\bs D$, the intersection would be private for the
other core of $\bs G$. It would then be covered by an edge of $H_D$,
contradicting the disjointness of the sets $\Lambda_D$.

\emph{A unique covering edge.} Let $\bs U\supseteq\bs G$ be priced. Each edge
covering $\bs U$ has its head in $U^*\subs G^*$. If such an edge $e$ is
retained at $\bs G$, its witness $\bs W$ is missed by the fixed edges and
therefore contains $\bs D$. Moreover $U^*\subs W^*$. Otherwise the argument
preceding~\eqref{internal} gives an edge of $Z_U\subs Z_G$ covering $\bs W$
that was already free before $\bs U$ was priced. This edge was present when
testing began at $\bs G$ and is tested after $e$ whether $e\in H\setminus Z$
or $e\in Z$. This contradicts the choice of $\bs W$. Until testing at
$\bs G$ ends there are only deletions, so each witness is still singly covered.
The witnesses of any two retained edges covering $\bs U$ contain $\bs D$
and have co-sets containing $U^*$. By the witness observation the two edges
coincide. Thus
$|\delta_Y(\bs U)|=1$ after testing at $\bs G$, and no later test changes
an edge with head in $U^*$.

An added edge
covering any priced $\bs T$ has its head in $T^*\cap G^*$ and its tail in
$T\setminus G$, so comparability forces $\bs G\subsetneq\bs T$.
Hence for a priced $\bs T$ containing no selected biset, $|\delta_Y(\bs T)|$
can only decrease. Its initial count satisfies~\eqref{counts} by~\eqref{q}
and the selection rule, and $\bs T$ remains covered. This proves the claim.

\paragraph*{Cost.}
By~\eqref{counts}, $|\delta_Y(\bs U)|\le g_{y_U}(s_U,t_U)$ for every priced
$\bs U$. Together with $r(Y)\le r(H)$ this gives
\[
 c(Y)=r(Y)+\sum_U\pi_U|\delta_Y(\bs U)|
 \le\Phi(H)\le\Phi(x)
 \le r^\top x+\tfrac54\sum_U\pi_U y_U
 \le\tfrac54 c^\top x.
\]
Here $g_{y_U}(a_U(x),b_U(x))\le5y_U/4$ bounds $\Phi(x)$.
The last inequality follows from $r\ge0$ and $c^\top x=r^\top x+\sum_U\pi_U y_U$.

\paragraph*{Proof of Theorem~\ref{thm:crossing} (directed ratio $13/4$).}
The preparation costs at most $2\tau(\FF)$. By Lemma~\ref{lem:two-cores},
$\FF^{J\cup I}$ has the local two-core property. Lemma~\ref{sharp} gives a
completion of cost at most $\tfrac54\tau(\FF^{J\cup I})\le\tfrac54\tau(\FF)$.
The total is at most $\tfrac{13}4\tau(\FF)$.\hfill$\square$

\section{Oracle implementation}\label{sec:implementation}
For a family with intersecting co-family, exact directed covering chooses an
inclusion-maximal uncovered member, raises its dual variable until a covering
edge is tight, and adds such an edge. Reverse deletion gives a cover whose cost
equals the LP value~\cite{FrankTardos1989,Frank2009,Nutov2014}. Reversing edges
and taking co-bisets gives the inclusion-minimal rule for intersecting families.

Feasibility is tested by querying all ordered pairs after installing all permitted
edges. The instance is feasible exactly when the residual family is empty.
For any subsequent edge set, the inclusion-minimal least answers give the residual cores. Keeping those with at most $p$ inner nodes gives the cores of the small residual family.
Thus a trial deletion in Lemma~\ref{lem:initial} is valid exactly when no residual
core lies in a member of $\PP$. Every maximal residual member $\bs A$ is the greatest
answer for any $u\in A$, $t\in A^*$. Thus, to find a maximal member containing
a given core, keep the greatest answers that contain it and take a maximal one.
For the pricing step of Lemma~\ref{sharp}, install the zero-cost edges and use
the same argument: among the greatest answers containing two cores of the
family at the start of completion, choose an inclusion-maximal one.

To restrict a query to the halo of $\bs C$, add the test edges $D\times D^*$ for
each other core $\bs D$. A member is covered by an edge of $D\times D^*$ exactly when it contains $\bs D$:
a member covered by such an edge crosses $\bs D$, so its intersection with
$\bs D$ equals $\bs D$ by minimality. Conversely, any member containing $\bs D$
is covered by a test edge from $D$ to its co-set. Querying $(u,v)$ now tests
whether the edge $uv$ covers a member of the halo. Each pricing step makes a new edge free, and both deletion
passes use polynomially many coverage tests. The prices have polynomial encoding length. Pricing uses only subtractions.
Each coefficient in $\Phi$ is a sum of terms formed from these prices,
the fixed denominators $y_U$, and loads that are either sums of input
coordinates or counts of chosen edges.

For a fixed ordered pair $u,t$, the members with $u\in A$ and $t\in A^*$
have an intersecting co-family. Exact directed covering optimizes over their
cover polytope with $0\le x\le1$: for general costs, fix all negative-cost
edges at one and cover the residual family. Optimization--separation equivalence
therefore gives a separation oracle for each pair. For $x\ge0$, replacing
each $x_e$ by $\min\{x_e,1\}$ preserves exactly the violated covering
inequalities. Testing all ordered pairs thus separates the covering
constraints of~\eqref{eq:augmentation-lp}.

For {\kca}, leave $u,t$ unsplit and split each internal node with a forward
edge of unit capacity and a reverse edge of infinite capacity. Base and
installed edges have infinite capacity. A minimum $u,t$ cut gives the pair oracle: a minimum value of $k$ yields the least
and greatest source sides, and a larger value gives an empty answer.
An installed edge from $u$ to $t$ also makes the answer empty.
For separation at $0\le x\le1$, give internal nodes capacity
$M>1+\sum_e x_e$, permitted edges capacity $x_e$, and base and installed
edges infinite capacity. For fixed $u,t$, a finite cut minimizes
$M|\partial\bs A|+x(\delta(\bs A))$ over bisets separating $u$ from $t$
and missed by the installed graph. Every such boundary has size at least
$k$, so a value below $Mk+1$ gives exactly a violated covering inequality
for that pair. This takes one maximum flow per ordered pair.

\paragraph*{Open problems.}
Can the directed ratio be reduced to $3$ for all $n,k$? In the undirected
case the preparation alone costs up to $4\tau(\FF)$ because it works with
both orientations. A direct undirected preparation or a lower bound on the
integrality gap would help determine how much room remains.

\paragraph*{AI disclosure.}
We used large language models to develop proof strategies, test arguments,
search for counterexamples, and revise the exposition. We check the proofs
ourselves and take full responsibility for the results and their presentation.

\begingroup
\small
\bibliography{references}
\endgroup
\clearpage
\appendix

\section{Proof of Lemma~\ref{lem:initial}}\label{sec:initialization}
We recall the primal--dual algorithm of~\cite[Theorem~2.11]{Nutov2014} in the form of~\cite[Section~6, Lemmas~14--15]{Nutov2018}.
The witness sets in part~(ii) come from the proof of Theorem~2.11,
specifically~\cite[Lemma~3.1]{Nutov2014}.
The forward run determines $\PP$. Each final member of $\PP$ has a unique
covering edge, whose tail lies in every surviving small core that meets that
member. We use this first to separate the sets $B_C$, then to show that each
has more than $p$ nodes.
Call an edge \textbf{dual-tight} when the sum of the dual variables of the
bisets it covers equals its cost.

\medskip
\begin{algorithm}[H]
\caption{Primal--dual initialization at threshold $p$}\label{alg:initial}
\KwInput{The family $\FF_{\le p}$ of Lemma~\ref{lem:initial} and permitted directed edges with costs.}
Set $J^+\gets\emptyset$, $\PP\gets\emptyset$, and all dual variables to zero\;
\While{$\FF^{J^+}_{\le p}$ is nonempty}{
Choose a core $\bs C$ of $\FF^{J^+}_{\le p}$\;
Set $\PP\gets\{\bs U\in\PP:\bs U\not\subseteq\bs C\}\cup\{\bs C\}$\;
Raise the dual variable of $\bs C$ until a covering edge is dual-tight, and add one such edge to $J^+$\;
}
Set $J\gets J^+$. Test the edges in reverse insertion order, deleting an edge if $J$ without it still covers every member of $\FF_{\le p}$ contained in some $\bs U\in\PP$\;
\Return{$(J,\PP)$}\;
\end{algorithm}
\medskip

Suppose a current member $\bs U$ of $\PP$ intersects a newly chosen $\bs C$.
Their intersection belongs to $\FF_{\le p}$. When $\bs U$ was chosen, both
$\bs U$ and $\bs C$ were uncovered. Their intersection was also uncovered,
since an edge covering it covers one of them. Minimality at that step gives
$\bs U\cap\bs C=\bs U$. Thus $\bs U\subs\bs C$, and the inner parts in $\PP$
stay disjoint. Every chosen core is contained in a final member of $\PP$.
For a final $\bs U$, no earlier edge covered it, and no later core meets $U$.
Its own edge $e(U)$ is therefore its unique covering edge in $J^+$ and must
be retained in $J$. This proves (i).

The family preserved by reverse deletion is intersecting. Indeed, two
intersecting members must lie in the same $\bs U\in\PP$, and their inner union has size at
most $p$. Every core chosen in the forward run belongs to this family and
was minimal among its uncovered members. Hence the run and its reverse deletion
are an execution of the exact primal--dual algorithm for an intersecting
family~\cite{Nutov2014}. Its cost is at most $\tau(\FF_{\le p})$.
Section~\ref{sec:implementation} states the exact covering rule.

Distinct small $\FF^J$-cores have disjoint inner parts, since an intersection
would be an uncovered member contained in both. If $\bs U\in\PP$ meets such a
core $\bs C$, their intersection belongs to $\FF_{\le p}$ and is contained in
$\bs U$. The edge of $J$ covering the intersection must cover $\bs U$, since
$J$ misses $\bs C$. Thus $e(U)$ has its tail in $C$. Each $\bs U$ meets at
most one of the small cores, proving disjointness of the sets $B_C$ in~\eqref{eq:witness}.

If $|B_C|\le p$, successively taking unions of $\bs C$ with the members of
$\PP$ meeting it gives a member of $\FF_{\le p}$ with inner part $B_C$, by
$p$-semi-intersection. Some edge of $J^+$ covers this union. Its tail lies in
a final member $\bs U$ of $\PP$, because every chosen core lies in one.
As the tail is in $B_C$ and the inner parts in $\PP$ are disjoint, $\bs U$
meets $\bs C$ and is one of the members in the union. The edge covers $\bs U$,
so it is $e(U)\in J$. Its tail belongs to $C$ and its head belongs to $C^*$,
contradicting that $\bs C$ is uncovered. This proves (ii).

\paragraph*{What can happen at $p=d/2$.}
For $h=d/2\ge3$, take disjoint sets $S=\{s_i\}_{i=1}^h$,
$T=\{t_i\}_{i=1}^h$, and $X_i,H_i$ of size $h-1$ for each $i$.
From the complete digraph delete $S\times T$ and, for every $i$,
$(\{s_i\}\cup X_i)\times(\{t_i\}\cup H_i)$; permit the deleted edges at unit cost.
The maximal pairs of nonempty shores with no base edge between them are
$(S,T)$ and, for each $i$,
$(\{s_i\},T\cup H_i)$, $(\{s_i\}\cup X_i,\{t_i\}\cup H_i)$, and
$(S\cup X_i,\{t_i\})$. Indeed, a shore using $X_i$ has all its heads in
$\{t_i\}\cup H_i$, and adding any $s_j$ with $j\ne i$ leaves only $t_i$.
All these pairs have total size $2h$, so the base connectivity is $n-2h$
and the list gives all tight bisets.
A legal run at $p=h$ buys first $s_i t_{i+1}$ at the singleton cores
(indices modulo $h$), then the already dual-tight edges $s_i z_i$, $z_i\in H_i$, at the cores with
inner parts $\{s_i\}\cup X_i$. Reverse deletion keeps exactly the latter
edges. Thus $B_S=S\cup\bigcup_i X_i$ has $h^2$ nodes, although $d=2h$.
Every surviving member contains the unique core with shores $(S,T)$, so
the next halo cover finishes the instance.

\section{Dependence on $d=n-k$}\label{app:parameter-bound}
\noindent\textbf{Proposition.} \emph{For directed graphs there is a
polynomial-time algorithm with LP-relative approximation ratio at most}
\[
 \rho(d)=
 \begin{cases}
  1, & d=2,3,\\[2pt]
  1+\dfrac54\cdot\dfrac{d-2}{d-1}, & 4\le d\le6,\\[5pt]
  1+\dfrac{d-2}{d-1}+\dfrac54\left(\dfrac{d-2}{d-1}\right)^2, & d\ge7.
 \end{cases}
\]
The bound increases to $13/4$. Since $\rho(14)=505/169$ and
$\rho(15)=2357/784$, it is below $3$ exactly for $2\le d\le14$.

Take an optimal extreme point $x$ of~\eqref{eq:augmentation-lp} with the
additional bounds $x_e\le1$. It is also optimal for the original LP, since
truncating any coordinate at one preserves coverage. The dependence on $d$
comes from the support of $x$ in an LP-tight covering inequality.

\paragraph*{Support at an LP-tight biset.}
Let $\bs S$ be a member of $\FF$ with $x(\delta(\bs S))=1$. The supported edges from $S$ to $S^*$ form
a bipartite forest. An edge with $x_e=1$ is the only supported edge in a
tight row containing it. Thus every edge on a supported cycle in this row has value
strictly between zero and one. On such a cycle one could perturb $x$ alternately
in both directions while preserving all tail and head sums in $S\times S^*$.
A member whose inner part misses $S$ or whose co-set misses $S^*$ is unaffected.
For $\bs A\subs\bs S$, the change in $x(\delta(\bs A))$ is the sum
of the changes at tails in $A$, since $S^*\subs A^*$. For $\bs S\subs\bs A$,
it is the sum of the changes at heads in $A^*$, since $S\subs A$. Both are zero. For every other member crossing $\bs S$, say $\bs A$, both
$\bs A\cap\bs S$ and $\bs A\cup\bs S$ belong to $\FF$ and are comparable
with $\bs S$.
For either perturbed vector $z$ we therefore have
\[
 z(\delta(\bs A))+1\ \ge\
 z(\delta(\bs A\cap\bs S))+z(\delta(\bs A\cup\bs S))\ \ge\ 2 .
\]
A sufficiently small perturbation respects the boxes, contradicting
extremality. Since $|S|+|S^*|=d$, at most $d-1$ supported edges cover an LP-tight biset.

For $d=2$ each member specifies one ordered pair. For $d=3$ each member has a
singleton inner part or a singleton co-set, and either determines the biset
by~\eqref{eq:inner-containment} applied to $\FF$ or $\FF^*$. Each column
therefore has at most one nonzero entry in each of the two row classes. The matrix
is a bipartite incidence matrix, with possible singleton columns, so the LP
is integral.

\paragraph*{Preparation at discounted prices.}
Put $\lambda=5/4$ and buy outright every edge with $x_e\ge1/\lambda$.
On the remaining edges write $t=x_e$. Use the original costs for completion.
The restriction of $x$ covers every residual family, since purchased edges
cover no surviving member.

For $d\ge7$, run the initialization of Lemma~\ref{lem:initial} on the residual
family at prices $c_e(1-t)(1-\lambda t)$, with $p=\lfloor(d-1)/3\rfloor$.
Then cover all residual halos by Lemma~\ref{lem:halo} at prices
$c_e(1-\lambda t)$. Lemma~\ref{lem:two-cores} gives the local two-core property.

For $4\le d\le6$, one purchase at prices $c_e(1-\lambda t)$ suffices.
For each surviving member with singleton inner part, buy a cheapest
covering edge. The singleton determines the member by~\eqref{eq:inner-containment},
so these requirements use disjoint tail domains. Every remaining core
has at least two inner nodes. If $\bs A,\bs B$ share a co-set node $v$,
the cores contained in either member all have $v$ in their co-sets, so their
inner parts are disjoint. These inner parts lie in $A\cup B$, which has
at most $d-1\le5$ nodes by~\eqref{eq:common-coset}. Hence at most two
such cores exist.

\paragraph*{Cost.}
In either case Lemma~\ref{sharp} completes the cover for at most
$\sum_{e\text{ remaining}}c_e\lambda x_e$. Work backwards through the
preparation. Let $R$ be the edges available before a purchase $P$ and
suppose the later purchases cost at most $\sum_{e\in R\setminus P}c_e a_e$.
At prices $c_e(1-a_e)$, the price of $P$ is at most
$\sum_{e\in R}c_e(1-a_e)x_e$, by the LP guarantee for that purchase:
Lemma~\ref{lem:initial}, Lemma~\ref{lem:halo}, or the disjoint singleton tails.
Thus
\[
 c(P)+\sum_{e\in R\setminus P}c_e a_e
 \le\sum_{e\in R}c_e\bigl[a_e+x_e(1-a_e)\bigr].
\]
With $j$ preparation purchases, the cost is therefore at most
$\sum_e c_e h_j(x_e)$, where
\[
 h_0(t)=\min\{1,\lambda t\},\qquad
 h_{j+1}(t)=h_j(t)+t\bigl(1-h_j(t)\bigr).
\]
For the outright purchases the allowance stays at one. For the remaining
edges, $1-h_0(t)=1-\lambda t$ and $1-h_1(t)=(1-t)(1-\lambda t)$
give exactly the prices above. Hence
$h_j(t)=1-(1-t)^j(1-\lambda t)_+$, where $(a)_+=\max\{a,0\}$.

For $j=1,2$, $h_j$ is concave. On $[0,1/\lambda]$ the product
$(1-t)^j(1-\lambda t)$ has nonnegative second derivative, since all factors
are nonnegative and their slopes have the same sign. At $1/\lambda$ the slope of $h_j$ drops to zero.
Let $y$ be an optimal dual of~\eqref{eq:augmentation-lp}. Complementary
slackness gives $c_e=\sum_{\bs S:e\in\delta(\bs S)}y_{\bs S}$ for $x_e>0$.
In every row with $y_{\bs S}>0$, the positive coordinates sum to one and
number at most $d-1$. Pad with zeros and apply Jensen's inequality:
\[
 \sum_e c_e h_j(x_e)
 =\sum_{\bs S}y_{\bs S}\sum_{e\in\delta(\bs S)}h_j(x_e)
 \le\tau(\FF)(d-1)h_j\bigl(1/(d-1)\bigr).
\]
Here $h_j(0)=0$ and $\sum_{\bs S}y_{\bs S}=\tau(\FF)$.
Substituting $j=1$ for $4\le d\le6$ and $j=2$ for $d\ge7$ gives
the two claimed bounds.\hfill$\square$

\end{document}